# Deep-learning assisted reduced order model for high-dimensional flow prediction from sparse data

Jiaxin Wu [a], Dunhui Xiao [b], Min Luo [a*]

[a] Ocean College, Zhejiang University, Zhoushan 316021, Zhejiang, China
[b] School of Mathematical Sciences, Tongji University, Shanghai 200092, China

## Abstract

The reconstruction and prediction of full-state flows from sparse data are of great scientific and engineering significance yet remain challenging, especially in applications where data are sparse and/or subjected to noise. To this end, this study proposes a deep-learning assisted non-intrusive reduced order model (named DCDMD) for high-dimensional flow prediction from sparse data. Based on the compressed sensing (CS)-Dynamic Mode Decomposition (DMD), the DCDMD model is distinguished by two novelties. Firstly, a sparse matrix is defined to overcome the strict random distribution condition of sensor locations in CS, thus allowing flexible sensor deployments and requiring very few sensors. Secondly, a deep-learning-based proxy is invoked to acquire coherent flow modes from the sparse data of high-dimensional flows, thereby addressing the issue of defining sparsity and the stringent incoherence condition in the conventional CSDMD. The two advantageous features, combined with the fact that the model retains flow physics in the online stage, lead to significant enhancements in accuracy and efficiency, as well as superior insensitivity to data noises (i.e., robustness), in both reconstruction and prediction of full-state flows. These are demonstrated by three benchmark examples, i.e., cylinder wake, weekly-mean sea surface temperature and isotropic turbulence in a periodic square area.

## 1. Introduction

Understanding flow patterns/structures is crucially essential in addressing many scientific and engineering problems, e.g., fluid turbulence [1], flow control [2, 3] and weather forecasting [4]. We have been witnessing rapid developments in measurement technologies and simulation methods that can produce plenty of data for analyzing relevant problems [5]. However, the massive data pose challenges to data storage and analysis efficiency [6-8]. Besides, since the data either from measurement or simulation are subjected to noises [9-11], the capturing of coherent physics without being affected by measurement noises remains another challenge. In this context, this work aims to develop an efficient and robust model capable of reconstructing and predicting global flows from highly-spare data with noises.

---

* Corresponding author.
 E-mail: min.luo@zju.edu.cn (M. Luo).

Fluid simulation or modeling approaches can be generally classified into three categories, i.e., the 'white-box', 'black-box' and 'gray-box' modeling [6, 12]. The 'white-box' model is based on the solving of defined governing equations like the Navier-Stokes equations, which is physically meaningful but can be computationally time-consuming. The 'black-box' model seeks an efficient mapping between inputs and specific outputs that substitutes for expensive equation solving or closure modeling [13-15]. Recently, deep learning has demonstrated its merits in building general correlations and has been getting more and more attention in fluid modeling [16-20]. However, the physical meaning or interpretability of results can be elusive, especially for problems with high nonlinearity and/or contaminated data. The 'gray-box' model is the third approach that incorporates the advantages of the 'white-box' and 'black-box' models. Specifically, it utilizes collected data with clarified physical frameworks to ensure both efficiency and interpretability in modeling.

One of the 'gray-box' approaches is to seek the low-order representation of a high-dimensional dynamic system, which is also termed the reduced order model (ROM). The rationality of this strategy is that although the full-state fluid dynamics contain a huge number of dimensionalities, the flow features are dominated by several low-order intrinsic modes (also called attractors or patterns) and hence can be approximated in a low-dimensional subspace. Therefore, ROMs are able to extract coherent flow modes that preserve the major physics at considerably low costs, with which the full-state flows can then be reconstructed and predicted.

Two canonical methods of ROMs are the proper orthogonal decomposition (POD) [21-23] and the dynamic mode decomposition (DMD) [24]. Both are capable of extracting low-dimensional dynamical features from snapshot data [25-27]. For POD, it identifies the spatially orthogonal modes sorted by energy (each of which contains multi-frequential components that affect the temporal evolution of fluid dynamics), and then reconstructs the full flow field by utilizing the first several modes that possess the majority of flow energy through linear superposition [28]. Note that the truncation of the low-energy modes may cause the missing of some frequential components that contribute significantly to temporal dynamics [29], adversely affecting future status prediction. In contrast, the modes extracted by DMD correspond to different frequencies and are sorted according to the dynamic importance. This feature, combined with the fact that the dynamic coefficients (each corresponding to a mode) naturally reflect the temporal dynamics, makes DMD a suitable approach for predicting fluid dynamics [25, 30].

The privilege of DMD makes it applied to various fluid dynamics problems such as cylinder wakes [31-33], jet flows [34], airfoil flows [35], cavity flows [36], and wall-bounded turbulence [37]. Recently, research efforts have been devoted to enhancing the DMD method from various aspects. For example, Chen et al. [38] introduced an optimized DMD for choosing modes and frequencies with lower residuals. Kutz et al. [39] developed a multi-resolution DMD to decompose the multiscale spatiotemporal features of the El Niño-southern oscillation more accurately. Baddoo et al. [40] developed a physics-informed DMD (piDMD) that incorporated prior knowledge to enhance the prediction of fluid flows. Scherl et al. [41] developed a robust DMD (RDMD) for separating outliners and noises from input data and hence improved the accuracy in dealing with noisy snapshot data. Note that almost all documented DMD applications in fluid dynamics are based on the premise of the availability of full-state data, yet in reality, only a fraction of the full-state data can be obtained due to the limitations in measurement devices or simulation capabilities [12]. This remains a challenge in applying DMD in real-life scenarios.

Compressed sensing (or compressive sampling, called CS for short) is a strategy that is capable of reconstructing full-state fluid fields from sparse data [42, 43]. It utilizes the sparsity and compressibility of signals in the frequency domain to inversely recover spatial-temporal dense signals by using heavily-compressed data. It has been applied in areas like image compression [44], computer vision [45], aerospace engineering [46], and medical imaging [47]. Fluid dynamics exhibit sparsity in the Fourier or Wavelet subspace, which enables the incorporation of CS with ROMs such as DMD and POD. Regarding DMD, Tu et al. [48] combined CS with DMD for the purpose of using lower sampling rates in handling moderate dynamic systems, and applied the approach in analyzing Particle Image Velocimetry (PIV) data. Brunton et al. [49] developed the compressed sensing DMD (CSDMD) and applied it to canonical examples like the sparsely-defined linear system and double gyre flow. Incorporating CS into DMD is a promising strategy to avoid the stringent premise of full-state data, yet the current CS-assisted DMD technologies still require massive data to obtain the desired accuracy.

Regarding the CS combined with POD, Brunton et al. [50] proposed a CS-POD approach with a library for classifying bifurcation parameters in nonlinear dynamical systems. The system parameters were successfully reconstructed/identified from sparse sensors, although the approach was computationally demanding and sensitive to noises. In the CS-POD work of Manohar et al. [51], the inaccuracy (due to random sampling) and the expensive solving process (e.g., convex optimization) of CS were demonstrated, and then an optimized sensor deployment strategy was proposed to partially address the limitations. In spite of the foregoing advancements, the accuracy, efficiency and robustness of CS-assisted ROMs need further developments such that its advantageous features in reconstructing full-state information from sparse data can be fully utilized.

In addition to the combination with ROMs, CS can also be integrated with Machine Learning (ML) strategies. For instance, Callaham et al. [5] introduced dictionary learning (through a pre-built library that contained prior knowledge and was costly) into POD to enhance the reconstruction accuracy from sparse data with noises. Erichson et al. [52] developed an end-to-end shallow neural network to reconstruct flow fields from sparse-sensor measurements, which showed satisfactory accuracy within the range of parameters covered by the training data. In terms of leveraging deep learning for CS, Fukami et al. [53] developed a Voronoi tessellation deep learning framework that optimally partitioned sensors to adaptive spatial structures for accommodating changes in sensor quantities and/or motions. This was indeed an end-to-end method that achieved full-state flow reconstruction from sparse measurements, and the authors pointed out the necessity of enforcing physical constraints. Zhang et al. [54] developed a CS-DMD framework that applied the Long Short Term Memory (LSTM) for predicting the sparse signals of dynamic flows and then invoked a Deep Neural Network (DNN) for full-state flow reconstruction. In short, the leverage of ML algorithms, more specifically supervised learning, for building end-to-end mapping has been demonstrated to be an effective strategy for advancing the capability of CS. Due to the lack of physical constraints, however, the convergence of supervised learning still faces challenges when the training data are sparse or compressed (which is inevitable in practical applications).

With the research background and knowledge gaps illustrated above, the present study proposes a Deep-learning enhanced Compressed-sensing DMD (DCDMD). The novelties of the model lie in two aspects: (1) a sparse matrix that overcomes the stringent random distribution condition of sensor locations in CS, thereby allowing flexible sensor deployments and requiring very few sensors; (2) a deep-learning assisted low-dimensional proxy that correlates sparse data and coherent flow modes, thus addressing the difficulty of defining data sparsity and the stringent incoherence condition in the conventional CSDMD. These features lead to superior accuracy, efficiency and robustness of DCDMD in identifying coherent flow modes, as well as in reconstructing and predicting complex global flows from very sparse data. In what follows, Section 2 introduces the fundamental theories of the conventional DMD and CS. Section 3 elaborates the concepts, formulations and numerical settings of the proposed DCDMD, as well as Section 4 describes the numerical implementations and settings. The performance of DCDMD is then demonstrated by three benchmark examples in Section 5, i.e., flow past a cylinder, global sea surface temperature, and isotropic turbulence in a periodic square domain. The research conclusions and findings are highlighted in Section 6.

## 2. Conventional CSDMD

### 2.1. Dynamic Mode Decomposition

Dynamic Mode Decomposition (DMD) is closely related to the Koopman analysis for seeking a linear operator for a high-dimensional system [55-57]. As a data-driven method, DMD extracts the spatiotemporal coherent modes or patterns of dynamic systems (e.g., fluid flows and structural deformations) from snapshot data [29, 36, 58]. For fluid dynamics problems, we can collect the flow field series (e.g., velocity, pressure, vorticity, temperature, concentration, etc.) with an evenly spaced time interval $\Delta t$, and pack them in a snapshot matrix $Z \in \mathbb{R}^{m \times n}$ as:

$$Z = \begin{bmatrix} | & | & & | & & | \\ x_1 & x_2 & ... & x_k & ... & x_n \\ | & | & & | & & | \end{bmatrix} \tag{1}$$

where $n$ denotes the number of time instants or snapshots, and $m$ indicates the number of measured points in each snapshot with $m \gg n$; $x_k \in \mathbb{R}^m$ is a column vector reshaped from the flow field snapshot at the $k$-th time instant.

DMD seeks the best-fit linear approximation of the operator $A$ that evolves the flow status from the current time instant to the next one via:

$$X' \approx AX \tag{2}$$

where $X = \begin{bmatrix} | & | & & | \\ x_1 & x_2 & \cdots & x_{n-1} \\ | & | & & | \end{bmatrix}$, and $X' = \begin{bmatrix} | & | & & | \\ x_2 & x_3 & \cdots & x_n \\ | & | & & | \end{bmatrix}$ are split from the entire data set.

The best-fit linear operator $A$ is constructed by minimizing the Frobenius norm of $X' - AX$ as:

$$A = \underset{A}{\arg\min} \left\| X' - AX \right\|_F \tag{3}$$

where $\| \cdot \|_F$ denotes the Frobenius norm. For computing $A$, applying the economy-size singular value decomposition (SVD) to $X$ and taking the first $r$ singular values (i.e., $r$ truncated modes) give:

$$X = U\Sigma V^* \tag{4}$$

where $\Sigma \in \mathbb{R}^{r \times r}$ is a diagonal matrix with the diagonal elements being the singular values; $U \in \mathbb{R}^{m \times r}$ and $V \in \mathbb{R}^{n \times r}$ represent the left and right singular vectors; the superscript * denotes the conjugate transpose, which gives that $U^* U = I$, $V^* V = I$. The operator $A$ can then be computed as

$$A = X' X^\dagger \tag{5}$$

where $X^\dagger = V\Sigma^{-1} U^*$ denotes the Moore-Penrose pseudoinverse. The eigenvalues of $A$ are the DMD eigenvalues.

For computational efficiency, the full matrix $A = U\tilde{A}U^*$ can be approximated by its low-dimensional representation $\tilde{A}$ ($\tilde{A}$ and $A$ have the same eigenvalues) as:

$$\tilde{A} = U^* X' V\Sigma^{-1} \tag{6}$$

A matrix $W$ that consists of the eigenvectors of $\tilde{A}$ is sought via taking the eigendecomposition of $\tilde{A}$:

$$\tilde{A}W = W\Lambda \tag{7}$$

where $\Lambda$ is a diagonal matrix containing DMD eigenvalues $\lambda_k$ (in the complex space) that correspond to the DMD operator $\tilde{A}$.

With $W$ clarified, the underlying DMD modes can then be constructed by projecting the left-hand side of Eq. (7) onto the $U$ space as [59]:

$$\Phi = X'V\Sigma^{-1}W \tag{8}$$

where each column $\phi_k$ of $\Phi$ corresponds to the $k$-th DMD eigenvector.

With the determined modes and eigenvalues, the full-state flow field $X$ can be reconstructed as:

$$X \approx \hat{X} = \Phi D_a V_{and} \tag{9}$$

where $D_a$ is diagonal matrix containing mode amplitudes $a_k = \Phi^\dagger x_1$; $V_{and}$ is the Vandermonde matrix consisting of DMD eigenvalues. Therefore, Eq. (9) can be written as:

$$\begin{bmatrix} | & | & & | \\ x_1 & x_2 & \cdots & x_{n-1} \\ | & | & & | \end{bmatrix} = \begin{bmatrix} | & | & & | \\ \phi_1 & \phi_2 & \cdots & \phi_r \\ | & | & & | \end{bmatrix} \begin{bmatrix} a_1 & & & \\ & a_2 & & \\ & & \ddots & \\ & & & a_r \end{bmatrix} \begin{bmatrix} 1 & \lambda_1 & \cdots & (\lambda_1)^{n-2} \\ 1 & \lambda_2 & \cdots & (\lambda_2)^{n-2} \\ \vdots & \vdots & & \vdots \\ 1 & \lambda_r & \cdots & (\lambda_r)^{n-2} \end{bmatrix} \tag{10}$$

With the above, snapshots at any future time $t$ can be approximated as:

$$\hat{X}(t) = \sum_{k=1}^{r} \phi_k \exp\left(\ln\frac{\lambda_k}{\Delta t} \cdot t\right) a_k \tag{11}$$

Letting $\omega_k = \ln\frac{\lambda_k}{\Delta t}$ (also in the complex space), Eq. (11) can be written as:

$$\hat{X}(t) = \sum_{k=1}^{r} \phi_k \exp(\omega_k t) a_k = \Phi \exp(\Omega t) D_a \tag{12}$$

where $\Omega = \text{diag}(\omega_k)$ is a diagonal matrix consisting of the complex eigenvalues $\omega_k$. The real part $\omega_r$ represents the growth/decay rate and the imaginary part $\omega_i$ indicates the temporal frequency [34]. The eigenvectors (i.e., DMD modes associated with $\omega_k$) also contain real and complex parts (i.e., $\phi_r$ and $\phi_i$).

Further simplifying Eq. (12) by letting $L_d = \exp(\Omega t) D_a$, which represents the DMD temporal dynamics of the flow field, we get:

$$\hat{X}(t) = \Phi L_d \tag{13}$$

This implies that the flow reconstruction and prediction in the DMD framework are essentially the computation of DMD modes and dynamics. However, the standard DMD approach requires the input of full-state data that is always inaccessible for large-scale and/or complex problems.

## 2.2. DMD with compressed sensing

Compressed sensing (CS) is a technology capable of recovering full-state flow information from sparse measurements. Given the fact that most natural signals, including fluid dynamics, are sparse in a certain space, the full-state data $X$ in Eq. (2) can be transformed as:

$$X = \Psi S \tag{14}$$

where $\Psi \in \mathbb{C}^{m \times m}$ is an appropriate basis (e.g., Fourier or Wavelet); $S = [s_1, s_2, ..., s_k, ..., s_{n-1}]$ contains the information of $X$ in basis $\Psi$ and each column $s_k \in \mathbb{R}^m$ has only $Q$ nonzero elements ($Q$ is known as the sparsity). That is to say, $X$ is $Q$-sparse in $\Psi$.

In applications where only a limited amount of data can be measured, the subsampled/compressed data $Y$ and the corresponding full-state data $X$ are related via a measurement or sparse matrix $C$ as:

$$Y = CX = C\Psi S \tag{15}$$

Similar to DMD, $X$ and $Y$ contain the information corresponding to the first to the $(n-1)$-th snapshots, with $Y = \begin{bmatrix} | & | & & | \\ y_1 & y_2 & \cdots & y_{n-1} \\ | & | & & | \end{bmatrix}$. $Y \in \mathbb{R}^{p \times (n-1)}$ has the same eigenvalues as $X$ because of the left invariance to the transform of $Y = CX$ [49]. $C \in \mathbb{R}^{p \times m}$ is a measurement matrix and $p$ ($Q < p \ll m$) indicates the number of sensors that should be in the order of $Q \log(m/Q)$ [60]. The subsampled and full-state information corresponding to the second to the $n$-th snapshots can be analyzed in a similar way. Given $Y$, $C$ and $\Psi$, we can solve Eq. (15) for $S$, with which the full-state information $X$ can be reconstructed by Eq. (14).

The above CS technology has been incorporated with DMD, leading to the so-called compressed sensing DMD (CSDMD) [49]. Instead of reconstructing coherent modes from full-state data in DMD, CSDMD enables the reconstruction of low-dimensional DMD modes from few measurement data. According to the invariance of DMD to unitary transformations, the DMD procedures introduced in Section 2.1 can be applied to the subsampled data matrix $Y$ and $Y'$, which yield the compressed modes $\varPhi_Y$ as [49]:

$$\varPhi_Y = C\varPhi_X = C\varPsi\varPhi_S \qquad (16)$$

where $\varPhi_Y$ contains DMD modes from the compressed data $Y$, and $\varPhi_S$ is the corresponding coefficient matrix. With $\varPhi_S$ and $\varPhi_X$ solved, the full-state information can be reconstructed via the linear superposition as that in DMD. This method possesses distinguishing advantages in relaxing the stringent data requirement and hence enhances computational efficiency.

An essential step of CSDMD is the determination of $\varPhi_S$. Since Eq. (16) is ill-conditioned and has infinite solutions, the sparsest solution $\hat{\varPhi}_S$ that approximates $\varPhi_S$ is sought by solving a convex $l_1$-minimization as:

$$\hat{\varPhi}_S = \underset{\varPhi_S}{\mathrm{argmin}} \left\| \varPhi_S \right\|_1, \text{ such that } \varPhi_Y = C\varPsi\varPhi_S \qquad (17)$$

where each column of $\hat{\varPhi}_S$ contains only $Q$ non-zero elements (i.e., sparsity $Q$) and $Q$ is user-defined in practice. Eq. (17) can be iteratively solved by convex optimization algorithms such as the compressed sensing matching pursuit (CoSaMP) [61], which is applied in the present study.

By incorporating DMD's capacity of field reconstruction through modes and CS's advantage of rebuilding modes from sparse measurements, CSDMD is of potential in recovering full-state flows from compressed data. However, the accuracy and/or efficiency of CSDMD remain to be further improved in dealing with real-life large-scale problems. The primary reason is that CS assumes the hydrodynamics being sparse in a certain space, while the sparsity $Q$ is always unknown beforehand in the chosen space. Another reason lies in the requirement of the measurement matrix $C$ being incoherent with the basis $\varPsi$ [49]. Theoretically, a random distribution scheme (e.g., Gaussian, Bernoulli, and sparse random [12]) can be adopted to assemble $C$. For implementing CSDMD, we adopt the scheme of deploying sensors whose locations satisfy the Gaussian random distribution. The number of sensors should be more than $Q\log(m/Q)$ and $10^2$, which is quite demanding [62]. Indeed, a sufficient number of measurement locations is also essential for reliably solving the convex optimization problem, i.e., Eq. (17). Practically, however, the above approach is expensive or even infeasible due to the cost and/or technical challenges in sensor deployment [12], as well as the heavy demand for computational resources.

## 3. Deep-learning assisted Compressed-sensing DMD

To enhance the practicality (in terms of deploying an affordable number of sensors at accessible locations) and efficiency of the conventional CSDMD, the present work proposes a Deep-learning assisted Compressed-sensing DMD (DCDMD) model that contains two distinct features, as elaborated below.

### 3.1. A sparse matrix compatible with flexible deployment and a small number of sensors

Building a subsampled data matrix $\hat{Y}$ is fundamental in CS. In the conventional CSDMD, the elements of the measurement matrix (i.e., $C$ as illustrated in Fig. 1 (a)) need to satisfy the random distribution, which may constrain practical applications because of the difficulties in deploying random sensors of a large quantity. To address this limitation, a measurement matrix that avoids the strict condition of random distribution and hence allows flexible sensor deployment is proposed.

The sparse measurement data $\hat{Y} \in \mathbb{R}^{m \times (n-1)}$ (to distinguish $Y$ in the conventional CSDMD) can be regarded as a filtered subset of the full-state data $X$ via:

$$\hat{Y} = \begin{bmatrix} | & | & & | & & | \\ \hat{y}_1 & \hat{y}_2 & \dots & \hat{y}_k & \dots & \hat{y}_{n-1} \\ | & | & & | & & | \end{bmatrix} = \left\langle C_s, \begin{bmatrix} | & | & & | & & | \\ x_1 & x_2 & \dots & x_k & \dots & x_{n-1} \\ | & | & & | & & | \end{bmatrix} \right\rangle = \left\langle C_s, X \right\rangle \tag{18}$$

where $C_s \in \mathbb{R}^{m \times (n-1)}$ is the sparse or measurement matrix of the proposed DCDMD. As illustrated in Fig. 1 (b), it is of the same shape as the data matrix $X$ ($Y$ or $\hat{Y}$) and each element of $C_s$ has a one-to-one correspondence to the field data. The element of $C_s$ in the $j$-th location at the $k$-th time instant, i.e., $c_{j,k}$, is one if a sensor exists at this location and zero otherwise. Although the sensor locations can vary with time, the present study considers spatially-fixed sensors that correspond to the scenario of deploying sensors at specific locations for monitoring certain physical quantities. In this case, the elements in a row of $C_s$, i.e., $c_{j,:}$, remain the same, as follows:

$$c_{j,:} = \begin{cases} 1, & j \subseteq \left\{ \xi \mid 0 < \xi \ll m, \xi \in \mathbb{Z} \right\} \\ 0, & j \not\subset \left\{ \xi \mid 0 < \xi \ll m, \xi \in \mathbb{Z} \right\} \end{cases} \tag{19}$$

where $\xi$ denotes the collection of locations with sensors.

The inner product of $C_s$ and $X$ is conducted, implying that only the data at locations with sensors are retained in the sparse data matrix (like a binary switch). For spatially fixed sensors, the filtered data matrix $\hat{Y}$ and the full-state data matrix $X$ in Eq. (18) can be written as:

$$\hat{Y} = \begin{bmatrix} \vdots & & \vdots & & \vdots \\ x_{a,1} & \dots & x_{a,k} & \dots & x_{a,n-1} \\ \vdots & & \vdots & & \vdots \\ 0 & \dots & 0 & \dots & 0 \\ \vdots & & \vdots & & \vdots \end{bmatrix} = \left\langle \begin{bmatrix} \vdots & & \vdots & & \vdots \\ 1 & \dots & 1 & \dots & 1 \\ \vdots & & \vdots & & \vdots \\ 0 & \dots & 0 & \dots & 0 \\ \vdots & & \vdots & & \vdots \end{bmatrix}, \begin{bmatrix} \vdots & & \vdots & & \vdots \\ x_{a,1} & \dots & x_{a,k} & \dots & x_{a,n-1} \\ \vdots & & \vdots & & \vdots \\ x_{b,1} & \dots & x_{b,k} & \dots & x_{b,n-1} \\ \vdots & & \vdots & & \vdots \end{bmatrix} \right\rangle \tag{20}$$

where elements $a \subseteq \{\xi \mid 0 < \xi \ll m, \xi \in \mathbb{Z}\}$ and $b \not\subset \{\xi \mid 0 < \xi \ll m, \xi \in \mathbb{Z}\}$ correspond to locations with and without sensors, respectively.

In this way, the filtered data matrix $\hat{Y}$ is highly sparse, and specifically, one deployed sensor corresponds to one row of nonzero elements. As will be demonstrated in Section 5, the proposed DCDMD requires very few sensors (much less than the sensor number required in CSDMD) and hence the requirement for data storage is much released. The fact that the locations (and thus indexes) of nonzero elements are fixed eases the implementation of the Compressed Sparse Row (CSR) storage technique, further enhancing the computational efficiency.

The proposed sparse matrix also avoids the strict Gaussian distribution condition of sensor locations (which is directly related to the incoherence condition in CSDMD). This offers great convenience in deploying sensors at optimized and accessible locations. In practical applications, exceptional situations like position offset and/or data interrupt (equivalent to missing sensors) are inevitable for some sensors, which change the values of the corresponding elements in the sparse matrix. In CSDMD, the variation of some sensor locations necessitates the resetting of the sparse matrix such that the Gaussian distribution condition is satisfied. This, in practice, implies the re-deployment of sensors, which can be costly and even infeasible. The present sparse matrix scheme overcomes these issues naturally.

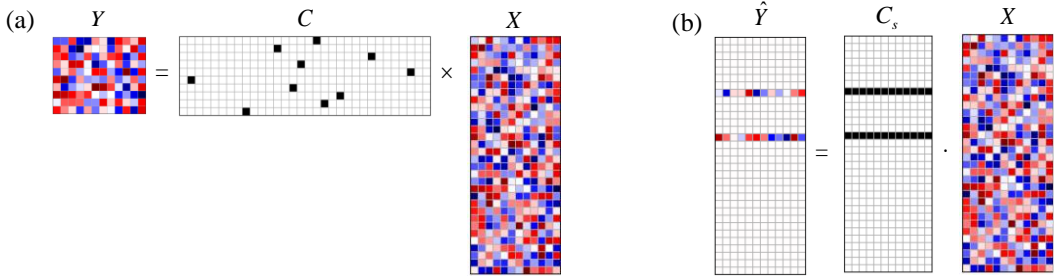

Fig. 1. Subsampling data from full-state snapshots by spatially-fixed sensors: (a) Conventional CS in which the sensor distribution needs to follow the random distribution [49]; (b) The present scheme that is compatible with flexibly-deployed and much-fewer sensors

### 3.2. A deep-learning strategy correlating sparse measurements and DMD modes

With the sparse measurement data matrix $\hat{Y}$ determined, the next essential step is identifying the DMD modes. To enhance the computational efficiency (an issue not fully resolved in conventional CSDMD), the present study invokes the deep learning strategy on building mapping relationships between sparse measurements and DMD models via a Deep neural network (DNN) proxy. The mapping function $\Gamma$ can be described as:

$$\hat{\Phi} = \Gamma\left(\hat{Y}\right) \tag{21}$$

where $\hat{\Phi}$ is a matrix approximating the DMD mode matrix $\Phi$; $\hat{Y}$ denotes the subsampled data collected from experiments or compressed from full-state numerical results.

The mapping function $\Gamma$ is trained by the supervised learning approach DNN in the offline stage, where the output is optimized to approach the DMD modes of the prior-knowledge information. As illustrated in Fig. 2, the DNN comprises an input layer receiving the compressed signals $\hat{Y}$, several hidden layers seeking an optimized correlation between inputs and outputs, and an output layer exporting the approximated modes $\hat{\Phi}$. For a DNN system with the sum of hidden and output layers being $L$ and each hidden layer containing $N$ neurons, the $j$-th neuron in the $l$-th layer (i.e., $\tau_j^{(l)}$) produces an output $h_j^{(l)}$ from the received signals (i.e., $h_k^{(l-1)}$) and a linear/nonlinear transformation (i.e., the activation function), as:

$$h_j^{(l)} = \gamma\left(\sum_{k=1}^{N} w_{jk}^{(l)} h_k^{(l-1)} + b_j^{(l)}\right), 1 \le l \le L, 1 \le j \le N \tag{22}$$

where $w_{jk}^{(l)}$ denotes the weight of the $j$-th neuron connecting to the $k$-th neuron in the previous layer; $b$ is a bias term; $\gamma$ is a user-defined activation function linearly/nonlinearly transforming the weighted inputs with biases to the next layer (note that no activation function is applied to the output layer).

The DNN architecture utilized in the present study is sketched in Fig. 2 and its hyperparameters are listed in Table. 1. Three hidden layers ($L = 4$) with 32 neurons in each layer ($N = 32$) are adopted for balancing accuracy (in terms of identifying the highly-nonlinear mapping) and efficiency (in terms of training time and computing resources). ReLU is selected as the activation function for its sufficient accuracy and suitable convergence property [63], which reads:

$$\gamma(z) = \max(0, z) \tag{23}$$

The adaptive moment estimation (ADAM) is chosen as the optimizer for updating the weight matrices $\boldsymbol{W}$ and bias vector $\boldsymbol{b}$, which minimizes the mean-squared error (MSE) loss $\mathbf{L}(\boldsymbol{W}, \boldsymbol{b})$ in the following form:

$$\underset{\boldsymbol{W}, \boldsymbol{b}}{\arg\min} \mathbf{L}(\boldsymbol{W}, \boldsymbol{b}) = \left\|\Phi - \Gamma\left(\hat{Y}\right)\right\|_2^2 \tag{24}$$

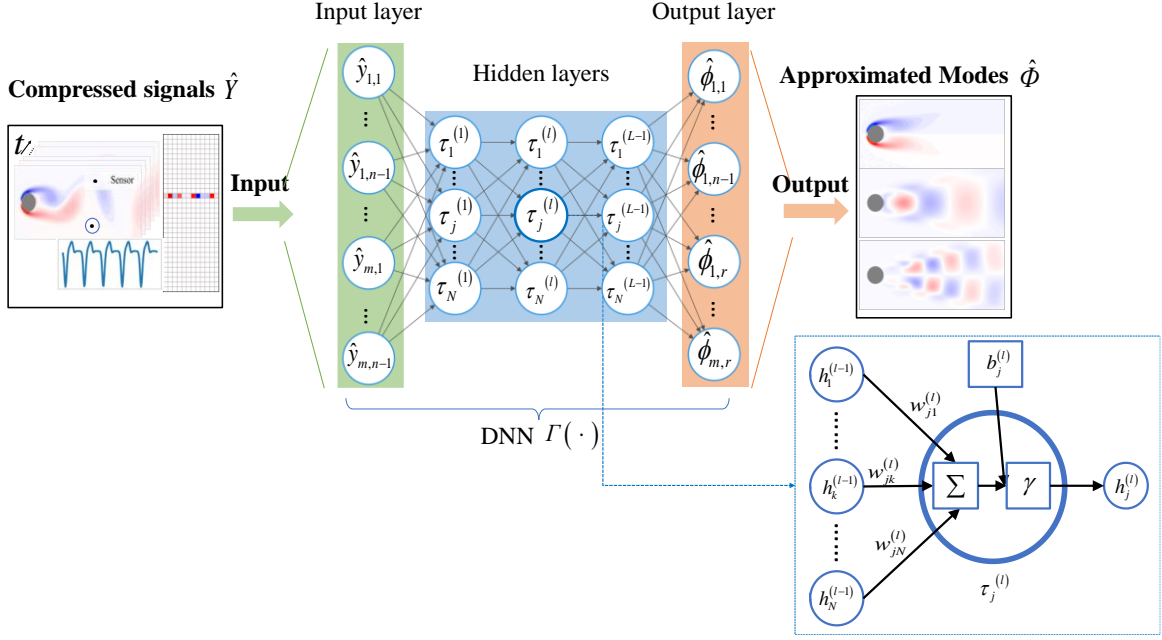

Fig. 2. Architecture of DNN for identifying the approximate DMD modes $\hat{\Phi}$ from sparse measurements $\hat{Y}$. The zoom-in plot in the right-bottom corner illustrates the structure of a single neuron.

Table. 1. Hyperparameters of the adopted DNN model

| Parameters | Settings | Parameters | Settings |
|---|---|---|---|
| **Number of hidden layers** | 3 | **Epochs** | 1000 |
| **Number of input layer neurons** | $m \times n$ | **Activation** | ReLu |
| **1st layer size** | 32 | **Loss function** | MSE |
| **2nd layer size** | 32 | **Optimizer** | ADAM |
| **3rd layer size** | 32 | **Learning rate** | 0.001 |
| **Number of output layer neurons** | $m \times r$ | **1st decay rate** | 0.9 |
| **Batch size** | 32 | **2nd decay rate** | 0.999 |

This approach avoids the difficulty of defining the sparsity $Q$ in the conventional CSDMD. With $\Gamma$ established, the DMD modes can be computed efficiently in the online stage, enabling fast and, ideally, real-time prediction. Additionally, this way of identifying the coherent modes avoids the essential assumption of the conventional CSDMD in that the measurement matrix $C$ should be incoherent with the basis $\Psi$.

With the approximate DMD models $\hat{\Phi}$ solved, the approximate spatiotemporal full-state flow fields $\tilde{X}(t)$ can be reconstructed/predicted by:

$$\hat{X}(t) \approx \tilde{X}(t) = \hat{\Phi} L_d \qquad (25)$$

where $L_d$ contains the dynamic coefficients corresponding to DMD modes, which are determined in the offline stage by analyzing the full-state data of a specific dynamic problem and pre-stored for recalling in the online stage of the same problem (as illustrated in Fig. 3). In this way, the prior knowledge, or more specifically, the dynamical importance of each mode, is retained in system reconstruction and prediction.

### 3.3. Sum-up notes of DCDMD

In summary, a sparse measurement matrix that allows flexible sensor deployments and requires very few sensors, as well as a DNN-based proxy that relates sparse measurements to DMD modes, are proposed for achieving enhanced practicality, efficiency and robustness. The enhanced model is termed the Deep-learning assisted Compressed-sensing DMD (DCDMD) with the key concepts/procedures illustrated in Fig. 3. The workflow can be divided into two parts, i.e., offline training and online forecasting. In the offline stage, the prior knowledge, i.e., coherent modes and dynamics, are extracted from the full-state data via DMD. Compressed/sparse data of the full-state fields can be obtained from a small number of flexibly-deployed sensors. By training a DNN, the mapping between the compressed data and the identified DMD modes is established. The coherent dynamics corresponding to DMD modes are stored in a library to be called online, which helps retain physics during reconstruction and prediction. In the online stage, the instantaneous sparse measurements are used as the input of the trained DNN to get the approximated DMD modes (i.e., $\hat{\Phi}$ in Eq. (25)) in a very fast manner, the production of which with the dynamics (i.e., $L_d$ in Eq. (25)) stored in the offline stage produces the full-state fields efficiently or even in real time.

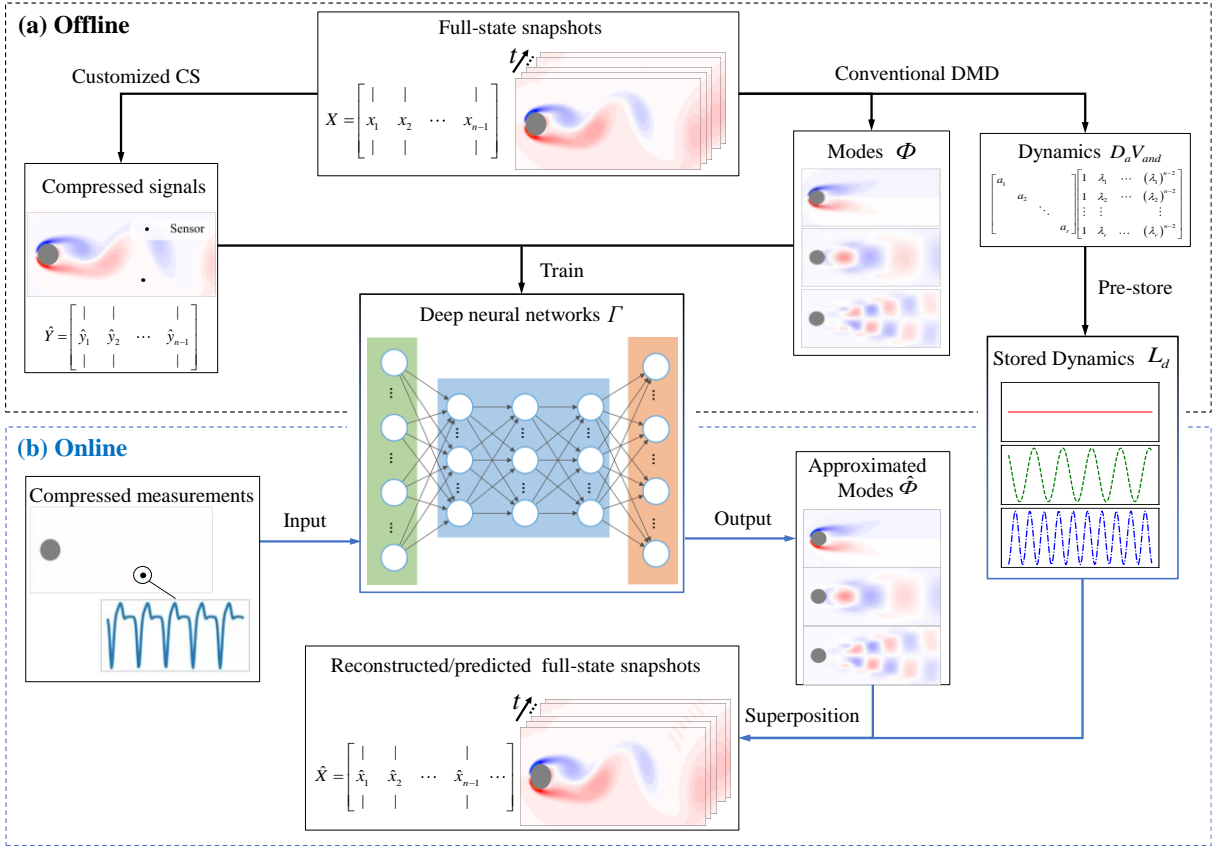

Fig. 3. Workflow of DCDMD in full-field reconstruction/prediction from sparse measurements: (a) offline training; (b) online reconstruction/prediction. Black dots indicate sensors.

## 4. Numerical implementations and settings

The code of DCDMD is developed based on the open-source package pyDMD, which executes the DMD method in the Python environment [64]. The DNN utilized in the present study is based on the Keras package applying the Tensorflow backend. Three numerical examples are studied and are presented in Section 5. All computations are conducted in a desktop equipped with an AMD Ryzen 5 3600 CPU and a NVIDIA RTX 3060 (8GB) GPU, with the offline training and online predicting of Deep Neural Networks executed on Tensorflow by GPU and the other computations/processes executed by CPU.

For demonstrating the superiority of DCDMD in accuracy and robustness (from the perspective of being insensitive to signal noises), three comparison groups are studied. They are: (1) DCDMD with one randomly-deployed spatially-fixed sensor, i.e., DCDMD with $p = 1$; (2) CSDMD with 1000 randomly-placed spatially-fixed sensors, i.e., CSDMD with $p = 1000$; and (3) full-state DMD (without CS). In CSDMD computations, the sparsity $Q$ values of CoSaMP in the three cases are selected to be 25, 100 and 50, respectively (more details are referred to Appendix 7.1).

Both interpolation inside the training set (i.e., reconstruction) and extrapolation outside the training set (i.e., prediction) are studied in the present study. The primary error quantification indicator is the root-mean-square error (RMSE), evaluated as follows:

$$\text{RMSE} = \sqrt{\frac{1}{n}\frac{1}{m}\sum_{i=1}^{m}\sum_{j=1}^{n}\left(x_{i,j} - \hat{x}_{i,j}\right)^2} \tag{26}$$

where $\hat{x}_{i,j}$ represents the predicted value and $x_{i,j}$ denotes the corresponding measured value. Note that Eq. (26) can be used for spatial- or temporal-averaged error estimation.

## 5. Results and discussion

Three numerical examples are studied to test the performance of the proposed DCDMD in constructing and predicting dynamic fluid flows.

### 5.1. Flow past a cylinder at Re = 100

The first case studied is the flow past a circular cylinder, as illustrated in Fig. 4. Full-state data are from the direct numerical simulation (DNS) based on the vorticity-stream function formulation:

$$\omega_t + \vec{u} \cdot \nabla \omega = \frac{1}{\text{Re}}\Delta\omega \tag{27}$$

where $\omega$ denotes the vorticity; $D$ is the diameter of the cylinder; $U_\infty$ is the free-stream velocity; Re = $DU_\infty / v = 100$, with $v$ being the kinematic viscosity. Eq. (27) is solved using the immersed boundary projection method [65, 66]. The computational domain is $9\,D \times 4\,D$, which contains $450 \times 200$ meshes; the computational time step is $\Delta t = 0.02$. The period of the vortex shedding in this case is $T = 6.0$ ($300\Delta t$). The data set contains 150 snapshots of vorticity field and are split into two parts: the first 125 snapshots ($t = 0$ - 25.0, around $4T$) used for training; the remaining 25 snapshots ($t = 25.0$ - 30.0, around $1T$) used for validation. For balancing computational efficiency and accuracy, the first 28 modes that contain more than 99.99% of the total dynamical information of the flow are adopted for reconstruction in this case.

### 5.1.1. Accuracy assessment

The accuracy of three comparison groups, i.e., DCDMD with $p = 1$, CSDMD with $p = 1000$, and DMD, in reconstructing fluid modes and constructing/predicting flow fields are compared. In DMD, the full-state data are used and the results serve as the benchmark. In CSDMD, 1000 sensors are randomly deployed in space and remain fixed in each run. In DCDMD, one sensor is randomly placed and keeps fixed (see Fig. 4). Five repeated runs of CSDMD and DCDMD with random sensor locations are conducted to avoid random errors caused by sensor deployments. It is found that results from the repeats are almost identical. Therefore, the results of one run are discussed in the following.

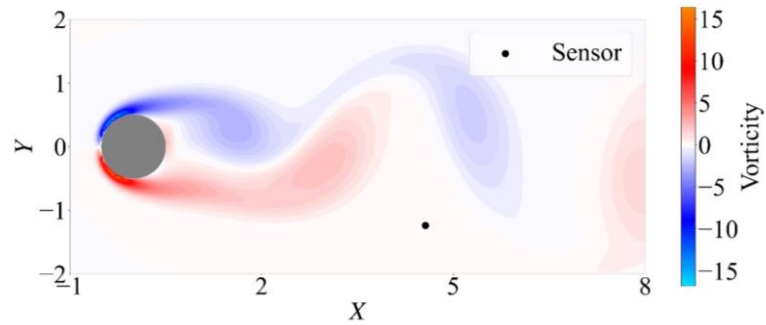

Fig. 4. Full-state data of flow past a circular cylinder and a randomly deployed sensor (black dot) in a DCDMD run

First, the DMD modes, with each representing the spatial structure of flows associated with the growth/decay rate at a fixed oscillation frequency, are analyzed. The real parts of the modes identified by DMD, CSDMD and the present DCDMD are shown in Fig. 6, and the corresponding imaginary parts are displayed in Fig. 6. The imaginary component of Mode 1 is zero, implying the stable mean flow without oscillation. Besides, Mode 1 has the most dynamical contribution. The modes from Mode 2 exhibit oscillating flow patterns (at fixed frequencies) as reflected by the contours of the imaginary components. Specifically, Modes 2 and 3, Modes 4 and 5, and so on, are conjugated mode pairs with the same real components and conjugated imaginary components (thus only Modes 3 and 5 are presented here). The real and imaginary parts of Mode 5 manifest asymmetric distributions along the longitudinal axis of the domain, which is closely associated with the Kármán vortex shedding downstream of the cylinder. Regarding CSDMD, although the overall trends of the modes captured are similar to those by DMD, the spatial distributions and amplitudes have evident discrepancies with the DMD results, which will further lead to errors in field reconstruction. In contrast, the modes identified by the present DCDMD from highly-sparse measurements (with only one sensor) are very close to those identified by the conventional DMD approach based on full-state data. This demonstrates the advantage of the proposed DCDMD in mode recovery.

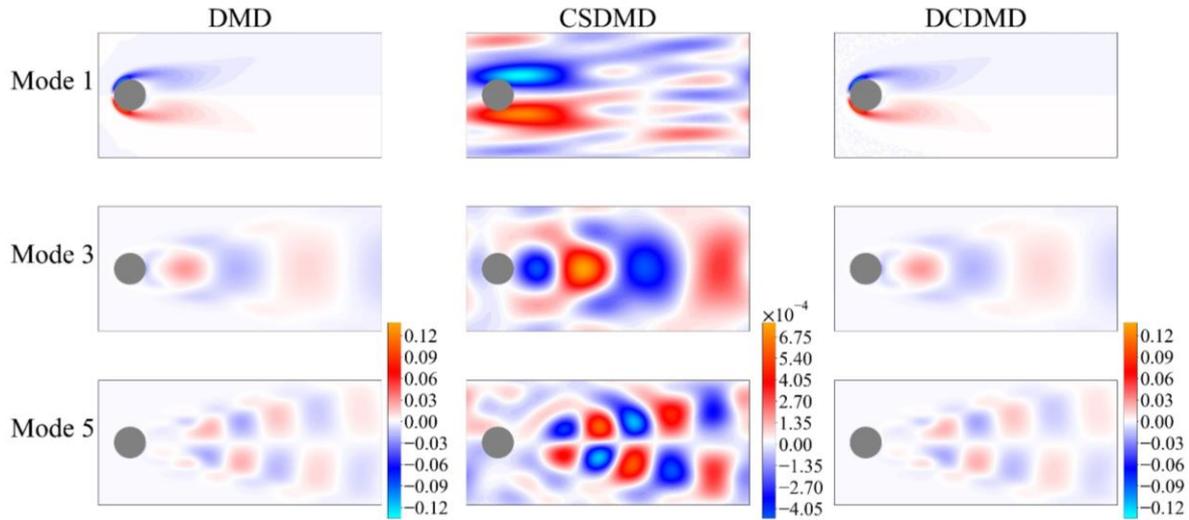

Fig. 5. Real parts of modes (cylinder wake) identified by DMD, CSDMD and DCDMD

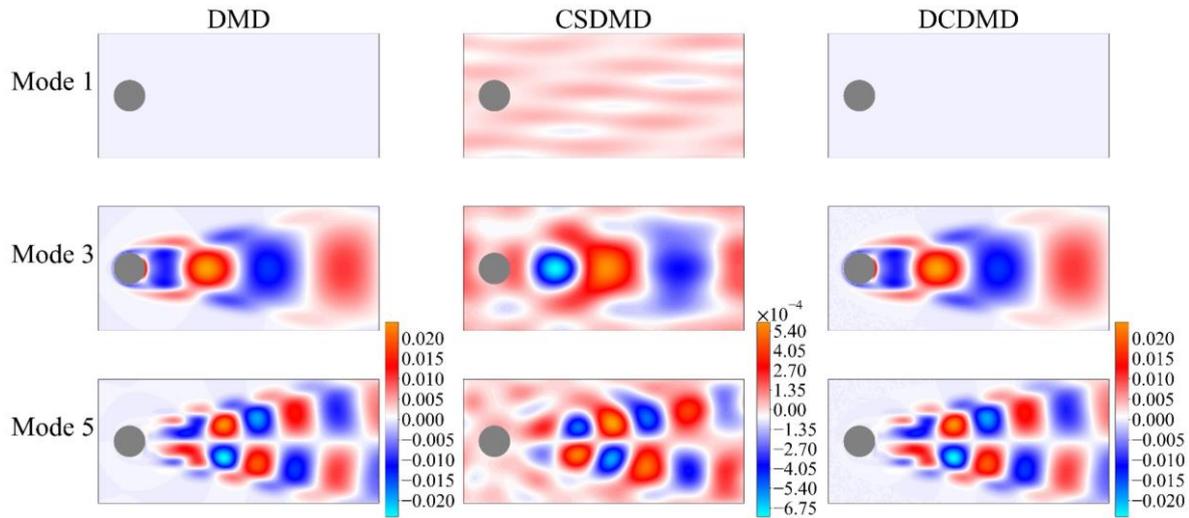

Fig. 6. Imaginary parts of modes (cylinder wake) identified by DMD, CSDMD and DCDMD

With the fluid modes identified, the full-state information can be efficiently computed by the superposition of modes and temporal dynamics. The reconstructed/predicted vorticity fields by DMD, CSDMD and DCDMD are presented in Fig. 7 in comparison with the reference data (snapshots of DNS). The DMD that requires full-state data reconstructs the full-state information accurately. The CSDMD utilizing sparse measurements from 1000 randomly placed sensors has produced the general trend of the periodic vortex pattern, but evident discrepancies exist in the area upstream of the cylinder and the vortex shedding area (due to the uncertainty of defining sparsity without prior knowledge). On the other hand, the vorticity fields reconstructed by the proposed DCDMD are very close to the DMD and reference results even in the prediction stage at $t$ = 30.0. This, combined with the fact that sparse measurements from only one sensor (at only 0.006% of those in DMD) are utilized, demonstrates the superior accuracy of DCDMD.

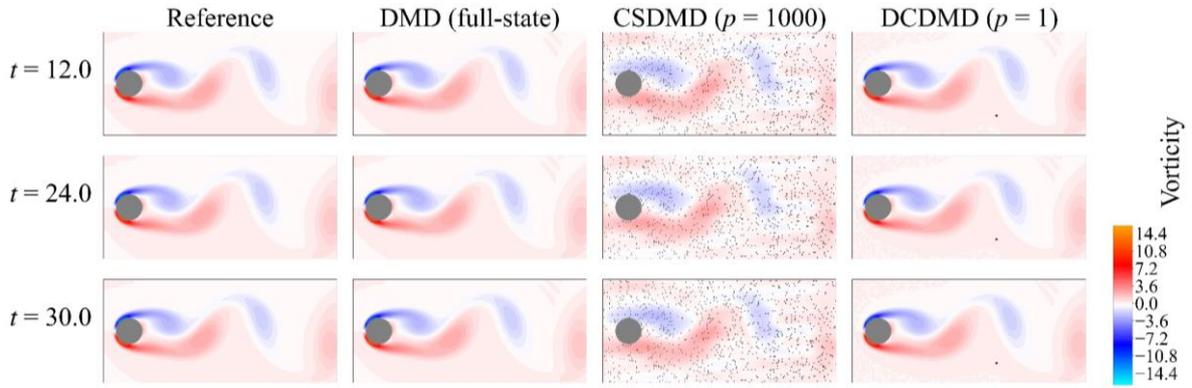

Fig. 7. Reconstructed vorticity fields at 12.0 ($2T$) and 24.0 ($4T$), and predicted vorticity fields at 30.0 ($5T$): (a) Reference data; (b) DMD; (c) CSDMD; (d) DCDMD. Black dots in the third and fourth rows indicate randomly deployed sensors.

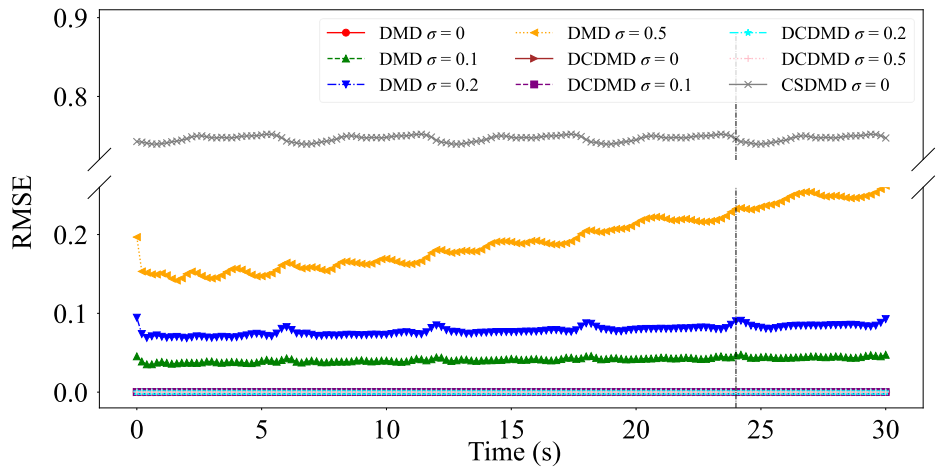

Fig. 8. Spatial-averaged RMSEs of DMD, CSDMD and DCDMD solutions subjected to different noise levels in the case of cylinder wake

The spatial-averaged RMSEs of the vorticity fields produced by DMD, CSDMD and DCDMD with respect to the reference values are evaluated and presented in Fig. 8 (the solid lines corresponding to $\sigma$ = 0). As can be seen, the RMSE of CSDMD results (around 0.75) is much larger than those of DCDMD and DMD solutions (with the peak value being around $1.4 \times 10^{-4}$, which is negligibly small). For DCDMD and DMD, the RMSEs are very close in both the reconstruction and prediction stages. Noteworthy is that the RMSEs in the prediction stage are at the same level (only 5.5% larger) as in the reconstruction stage, as further confirmed by the spatial-temporal averaged RMSE in Table. 2 (the row of $\sigma = 0$). This demonstrates the superior capability of the proposed DCDMD in predicting future events through extrapolating available datasets.

Table. 2. Spatial-temporal averaged RMSEs of DMD, CSDMD and DCDMD solutions subjected to different noise levels in the case of cylinder wake

| | $\sigma$ | DMD (full field) | | CSDMD (1000 sensors) | | DCDMD (1 sensor) | |
| | | Reconstruction | Prediction | Reconstruction | Prediction | Reconstruction | Prediction |
|---|---|---|---|---|---|---|---|
| RMSE | 0 | $1.08\times10^{-4}$ | $1.15\times10^{-4}$ | 0.75 | 0.75 | $1.09\times10^{-4}$ | $1.15\times10^{-4}$ |
| RMSE | 0.1 | $4.06\times10^{-2}$ | $4.48\times10^{-2}$ | \ | \ | $1.09\times10^{-4}$ | $1.15\times10^{-4}$ |
| RMSE | 0.2 | $7.62\times10^{-2}$ | $8.48\times10^{-2}$ | \ | \ | $1.09\times10^{-4}$ | $1.15\times10^{-4}$ |
| RMSE | 0.5 | 0.18 | 0.25 | \ | \ | $1.09\times10^{-4}$ | $1.15\times10^{-4}$ |

### 5.1.2. Robustness assessment

The sensitivity of the developed DCDMD to noise is studied by handling snapshot data with Gaussian white noises. Three noise-to-signal ratios, i.e., $\sigma = 0.1$, 0.2 and 0.5, are considered and the corresponding noises are added into the clean vorticity fields. Only results by DMD and DCDMD are presented because of their comparable performances on clean data. As shown in the middle column of Fig. 9, the vorticity fields reconstructed by DMD in different noise cases manifest differences. Besides, as depicted in Fig. 8, the RMSEs increase with time and the increase ratio is significantly larger for the case with higher noise ratios, which would adversely affect the prediction of nonlinear fluid dynamics. As listed in Table. 2, the RMSEs for $\sigma = 0$, 0.1, 0.2 and 0.5 are $1.08\times10^{-4}$, $4.06\times10^{-2}$, $7.62\times10^{-2}$ and 0.18, respectively, corresponding to an increase around two orders of magnitude. All these show DMD's sensitivity to noises and hence strong dependency on data quality. In contrast, the vorticity fields reconstructed by DCDMD are almost identical in different noise levels (see Fig. 9). Specifically, the RMSEs in the four cases (i.e., $\sigma = 0$, 0.1, 0.2 and 0.5) are all around $1.09\times10^{-4}$ (see Table. 2) and do not amplify in the time domain (see Fig. 8), which demonstrate the insensitivity of DCDMD to data noises. This advantageous feature is attributed to the fact that the DNN supervised learning successfully identifies the coherent flow modes, which, combined with the pre-stored dynamics of the flow, leads to accurate flow field reconstruction.

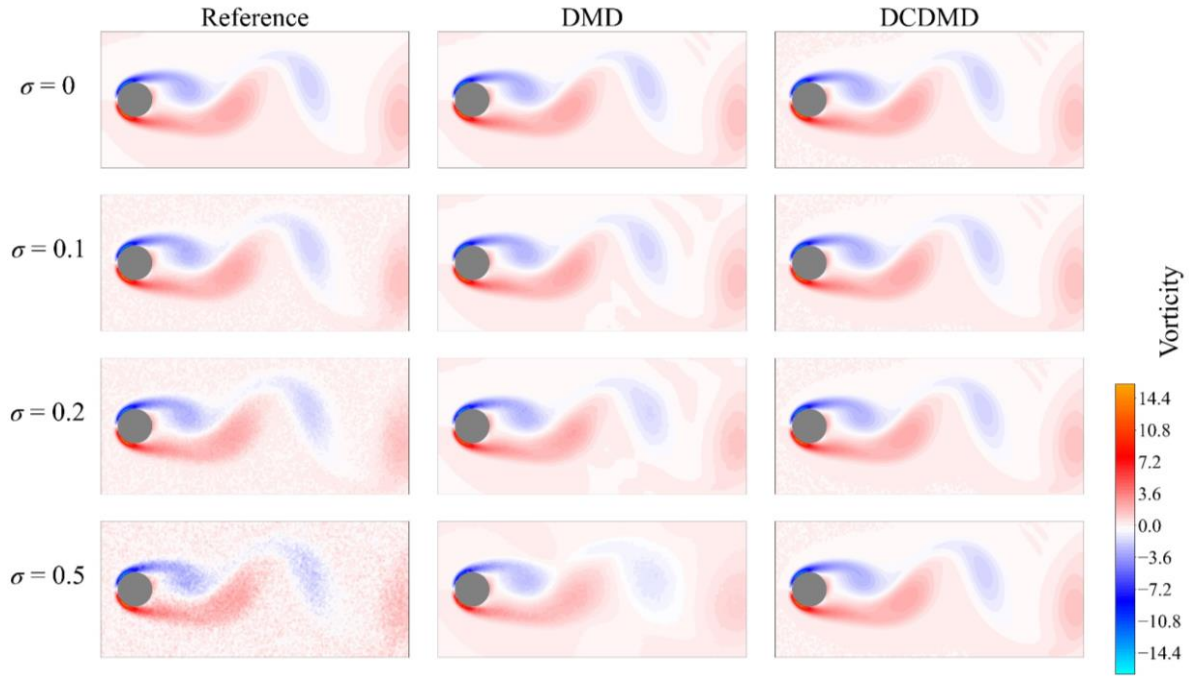

Fig. 9. Predicted vorticity fields at 30.0 ($5T$) with noise-to-signal ratios $\sigma = 0$, 0.1, 0.2 and 0.5:

(a) Reference; (b) DMD; (c) DCDMD

## 5.2. Weekly-mean sea surface temperature

The second example is the global weekly-mean sea surface temperature (SST) as illustrated in Fig. 10. In the accessible dataset published by the National Oceanic & Atmospheric Administration (NOAA), USA, a grid resolution of 1° in both latitude and longitude is applied onto the earth surface, amounting to $180 \times 360$ grids [67]. The temperature data on the grids are based on in situ/satellite measurements and numerical simulations. In our simulation, half of the resolution, which corresponds to $90 \times 180$ grids, is considered for the sake of computational efficiency. 314 snapshots with a time interval of one week from Year 1970 - 1976 are considered. Among them, the first 257 snapshots are used for training and the remaining 57 are used for validation. Compared to the previous case, the SST case involves more low-frequential dynamics; hence, no mode rank truncation is applied in field reconstruction. In the following, the results from one run of five repeats (results are almost identical) are presented.

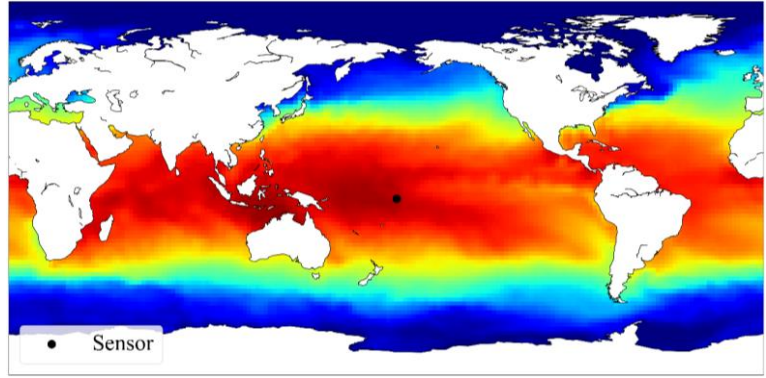

Fig. 10. Full-state data of global SST during the first week in Year 1970 and a randomly deployed sensor (black dot)

### 5.2.1. Accuracy assessment

The real and imaginary components of the coherent modes identified by the full-state DMD, CSDMD with $p = 1000$ and DCDMD with $p = 1$ are plotted in Fig. 11 and Fig. 12. The first mode identified manifests smoothly-descending temperature from the equator to poles, which represents the mean sea surface temperature. Although the solutions by the three models all show this trend, the pressure distributions identified by DMD and DCDMD are pretty close, while that by CSDMD shows higher temperatures near the equator region. Regarding Modes 3 and 5, which correspond to spatial oscillatory patterns, the deviations between CSDMD solutions and DMD and DCDMD ones (which are very close) are more extensive. This is attributed to the inaccuracy of CSDMD in iteratively approaching the global field from a few measurements and is related to the need for prior knowledge about sparsity in dynamics. In contrast, the limitations, e.g., tuning the sparsity unknown beforehand and repeating iterative computations, have been resolved by DCDMD via deep neural networks. With these advantages, the DCDMD identifies the features corresponding to the El Niño southern oscillation (ENSO) phenomenon in the eastern South Pacific near Peru's coast using data from only one sensor.

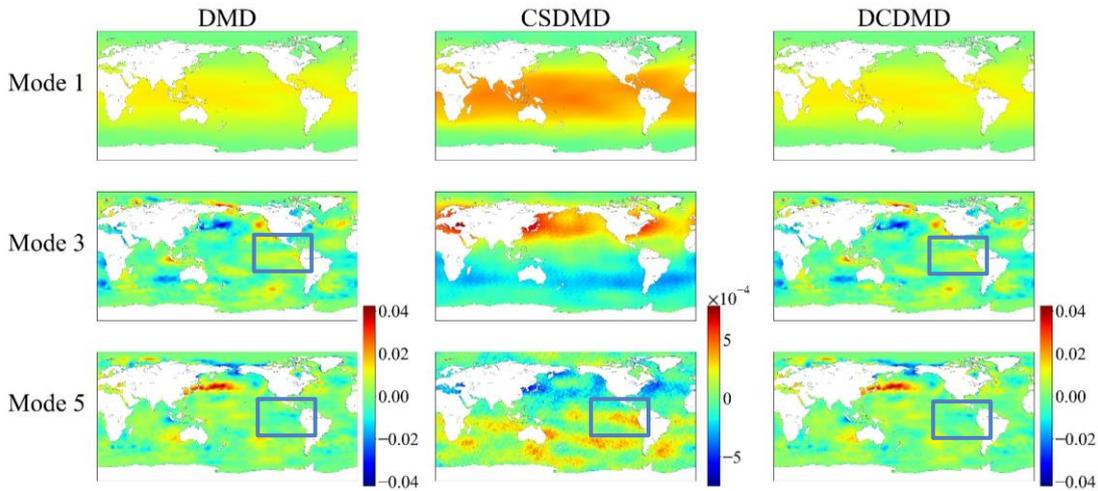

Fig. 11. Real parts of modes (global SST) identified by DMD, CSDMD and DCDMD (blue rectangles represent the ENSO region)

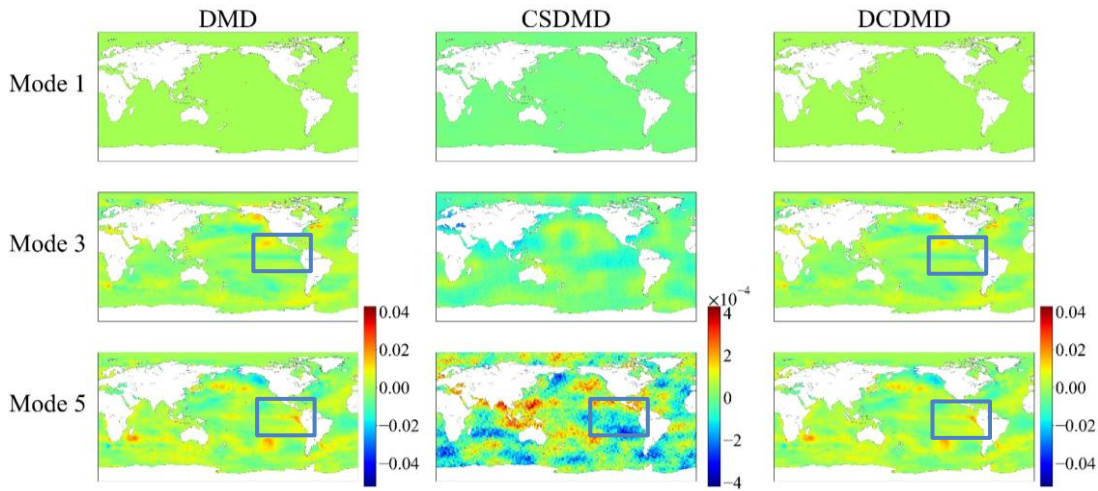

Fig. 12. Imaginary parts of modes (global SST) identified by DMD, CSDMD and DCDMD (blue rectangles represent the ENSO region)

Superposing the identified modes through dynamic coefficients, the SST fields can be constructed by the preceding three methods and are compared with the original fields (reference) in Fig. 13. Similar to the solutions of modes, the SST fields reconstructed by DCDMD with only one sensor match those by the full-state DMD and both are very close to the reference. In contrast, the CSDMD with 1000 sensors shows discrepancies, especially in areas where the SST exhibits strong nonlinearity (e.g., the estuary and ENSO regions, as shown in Fig. 14). The satisfactory accuracy of DCDMD using only one sensor results from the well-identified modes and the retained prior knowledge of dynamics. Notably, the ENSO phenomenon in the West of South America is successfully reconstructed and predicted by DCDMD, whereas the CSDMD solutions (by using one thousand sensors) show evident deviations from the reference. These demonstrate DCDMD's potential in extrapolating in space with measurements at a few locations, also known as the super-resolution reconstruction.

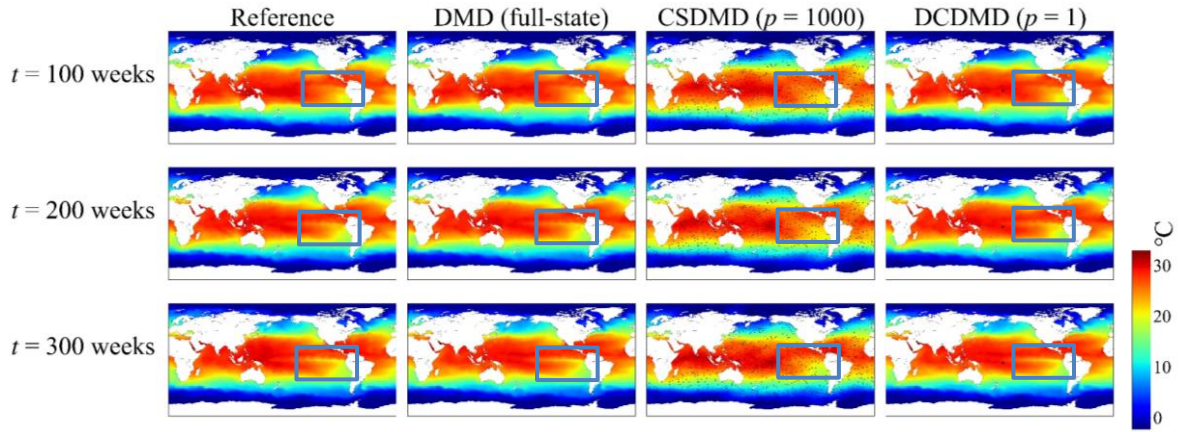

Fig. 13. Reconstructed SST fields at 100 weeks and 200 weeks, and predicted SST fields at 300 weeks: (a) Reference; (b) DMD; (c) CSDMD; (d) DCDMD. Black dots in the third and fourth rows indicate randomly deployed sensors. Blue rectangles represent the ENSO region.

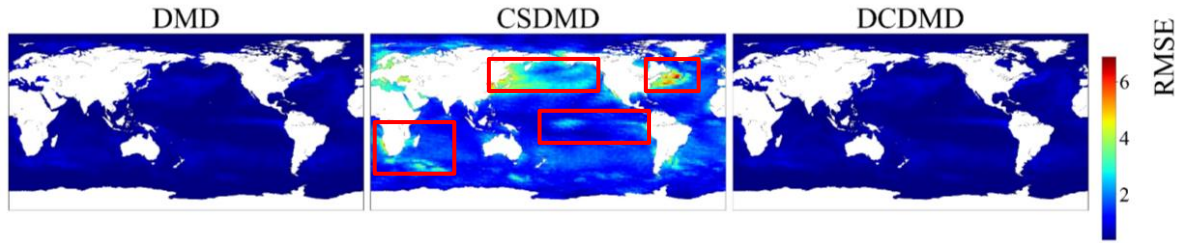

Fig. 14. Temporal-averaged RMSEs of the reconstructed SST distributions by DMD, CSDMD and DCDMD. Red rectangles correspond to the estuary and ENSO regions where the SST exhibits strong nonlinearity.

The spatial-averaged RMSEs of the three sets of solutions in the condition without data noises are evaluated and presented in Fig. 15 (see the solid lines for $\sigma = 0$). Generally, the RMSEs of DCDMD are very close to those of DMD (within a 1% difference), while those of CSDMD are much higher. Specifically, the RMSEs of CSDMD oscillate around the average value of 1.8 in both the reconstruction and prediction stages. The RMSEs of DCDMD in the reconstruction stage are around 0.4 and almost double in the prediction stage. These are corroborated by the spatial-temporal averaged errors in the first row of Table. 3. Such an increase in prediction errors is because the global SST convection contains multi-scale and strong-nonlinear flow patterns (e.g., ENSO), which are hard to describe perfectly by a linear dynamic system, as the DMD method assumes.

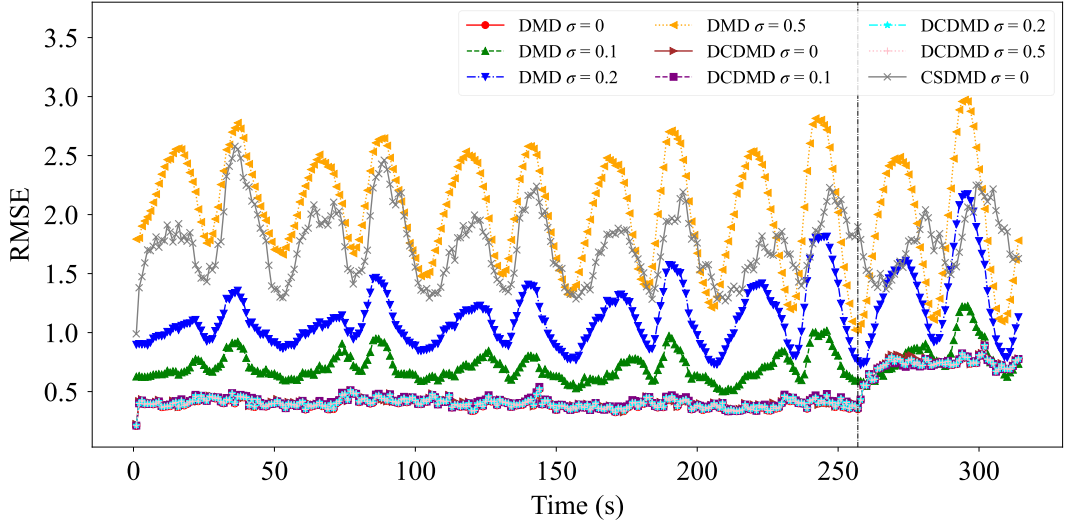

Fig. 15. Spatial-averaged RMSEs of DMD, CSDMD and DCDMD solutions subjected to different noise levels in the case of global SST

Table. 3. Spatial-temporal averaged RMSEs of DMD, CSDMD and DCDMD solutions subjected to different noise levels in the case of global SST

| | $\sigma$ | DMD (full-state) | | CSDMD (1000 sensors) | | DCDMD (1 sensor) | |
|---|---|---|---|---|---|---|---|
| | | Reconstruction | Prediction | Reconstruction | Prediction | Reconstruction | Prediction |
| **RMSE** | 0 | 0.40 | 0.73 | 1.81 | 1.76 | 0.41 | 0.74 |
| **RMSE** | 0.1 | 0.71 | 0.83 | \ | \ | 0.41 | 0.73 |
| **RMSE** | 0.2 | 1.12 | 1.39 | \ | \ | 0.40 | 0.73 |
| **RMSE** | 0.5 | 2.10 | 2.00 | \ | \ | 0.40 | 0.73 |

### 5.2.2. Robustness assessment

The predicted SST contours based on the snapshot data with Gaussian white noises are compared in Fig. 16. Similar to the case of flow past cylinder, the predicted fields by the traditional DMD vary with noise ratios. This trend can also be confirmed in the spatial-averaged error diagram in Fig. 15, where the RMSEs for $\sigma$ = 0, 0.1, 0.2 and 0.5 are 0.40, 0.71, 1.12 and 2.10, respectively, corresponding to a five-fold increase. In contrast, the DCDMD solutions at different noise levels are almost the same as qualitatively illustrated in Fig. 16. Quantitatively, the errors of the four noise cases stay at around 0.41 in reconstruction and about 0.74 in prediction (see Fig. 15 and Table. 3). This further demonstrates the superior robustness of DCDMD in terms of being insensitivity to data noises.

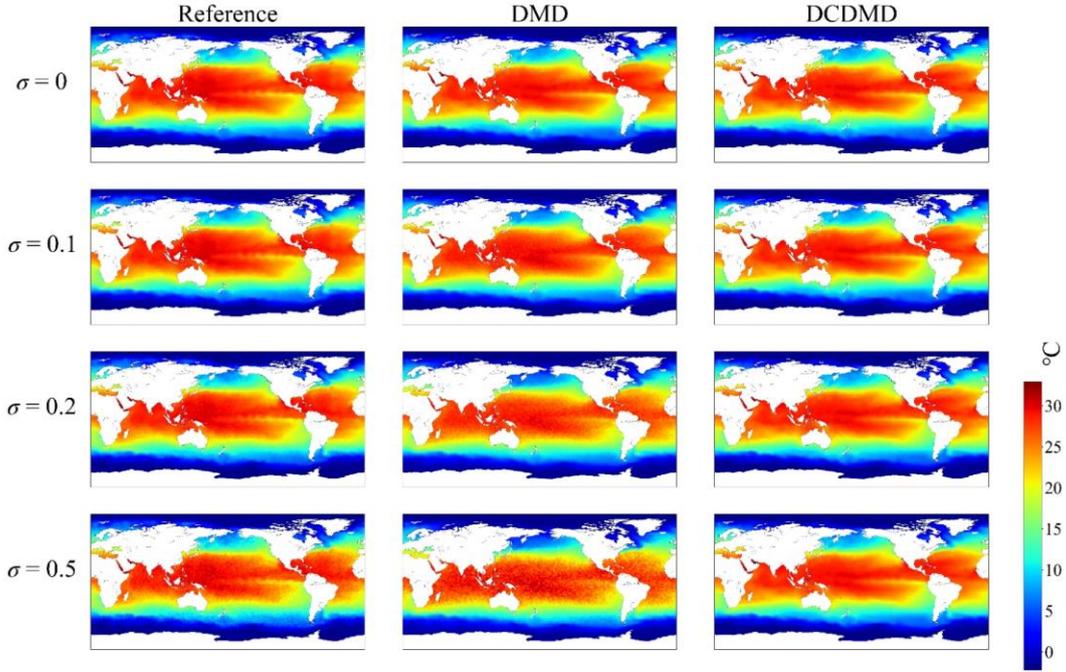

Fig. 16. Predicted SST fields at 300 weeks with noise-to-signal ratios of $\sigma = 0$, 0.1, 0.2 and 0.5: (a) Reference; (b) DMD; (c) DCDMD

## 5.3. Isotropic turbulence

The last example is the isotropic turbulence in a square area with periodic boundaries, which is a challenging problem in the fluid dynamics community. The reference data are from a DNS published in the Johns Hopkins Turbulence Database [68], in which the Taylor-scale Reynolds number $Re_\lambda = 433$ and the Taylor microscale $\lambda = 0.118$. The DNS was conducted in a three-dimensional computational domain of $[0, 2\pi]^3$ with $1024^3$ grids and a fixed time step of $\Delta t = 2 \times 10^{-4}$ s. In the present study, two-dimensional velocity data on the $x$-$y$ plane with $z$ coordinate being zero are extracted through the Lagrange-6 interpolation, which corresponds to $128 \times 128$ grids. 200 snapshots with a time interval of 0.02 s ($100\Delta t$) are considered, in which 150 snapshots are used for training (0.4 - 0.7 s) and the rest 50 are used for validation (0.7 - 0.8 s). No mode truncation is applied. The results from one run of five independent repeats are presented with one randomly deployed sensor illustrated in Fig. 17.

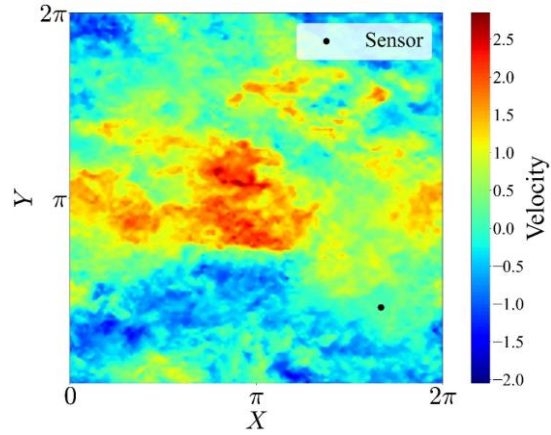

Fig. 17. Full-state data of isotropic turbulence in a square area with periodic boundaries and a randomly deployed sensor (black dot)

### 5.3.1. Accuracy assessment

The real and imaginary components of three representative modes are plotted in Fig. 18 and Fig. 19. In general, the modes identified by DCDMD utilizing a single sensor are very close to those by the full-state DMD, while those by CSDMD show large variations, especially for Modes 3 and 5. In the solutions produced by DMD/DCDMD, Mode 1 manifests a flow pattern associated with the mean velocity, where larger velocities exist in the upper middle of the domain. In Modes 3 and 5, irregularly distributed and fluctuated velocities are observed, indicating the chaotic features of the turbulent flow.

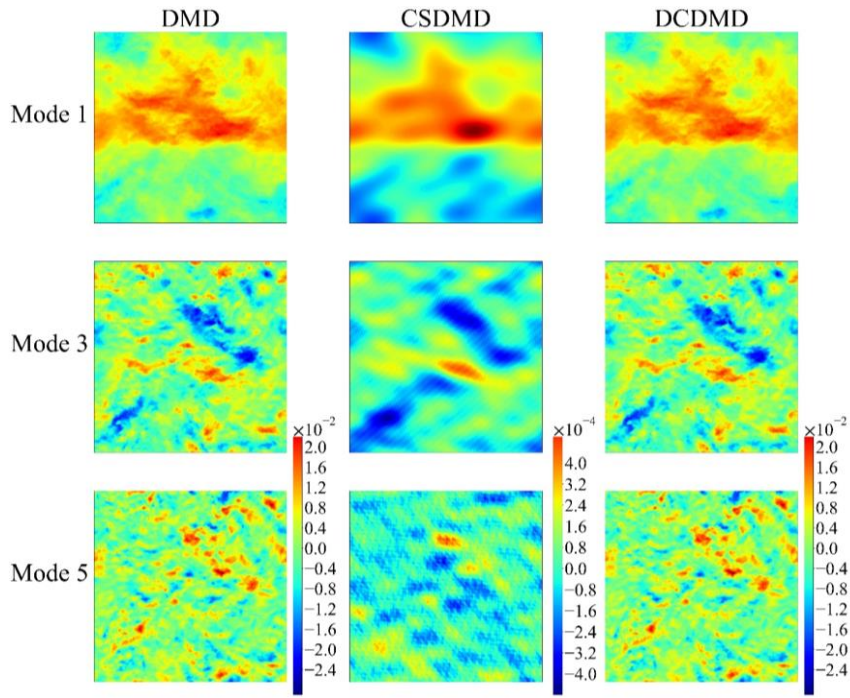

Fig. 18. Real parts of modes (isotropic turbulence) identified by DMD, CSDMD and DCDMD

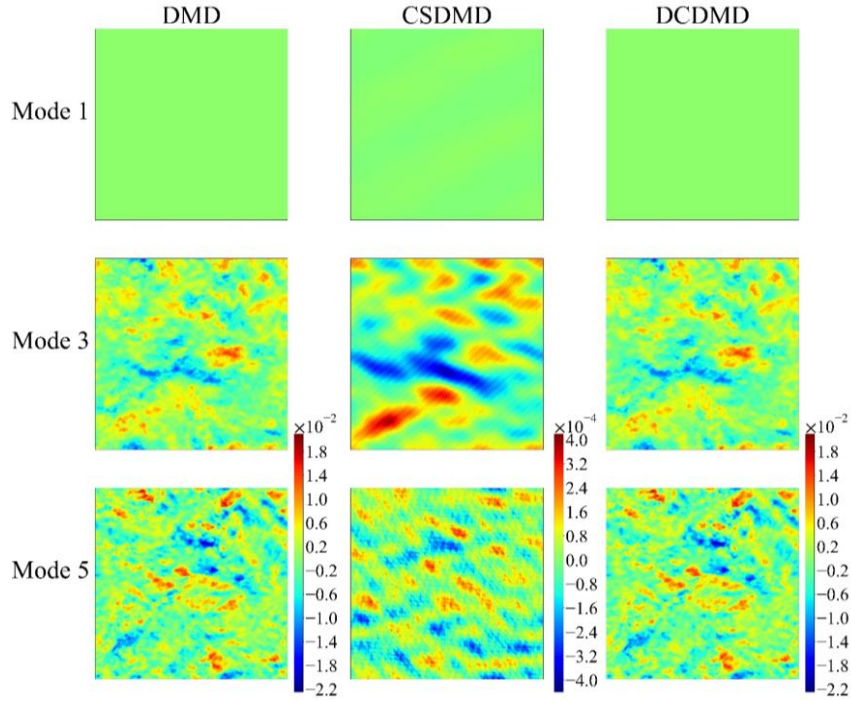

Fig. 19. Imaginary parts of modes (isotropic turbulence) identified by DMD, CSDMD and DCDMD

The velocity fields reconstructed through linear superposition of modes are presented in Fig. 20. Consistent with the foregoing two cases, DCDMD from sparse sensors produces almost identical results to DMD, while the CSDMD results based on 1000 sensors show evident differences, especially in areas with large velocities. This trend is further illustrated by the temporal-averaged RMSEs in Fig. 20. In the reconstruction stage, the spatial-averaged RMSE is around 0.06 for DMD/DCDMD and that for CSDMD is around 0.34 (see the solid lines in Fig. 21). In the prediction stage, the errors of DMD/DCDMD increase significantly (being around 50 times in the studied time range) and approach to that of the CSDMD result. Note that the errors are more significant in the regions with large velocities, as shown in Fig. 22. Compared to the previous two cases, fluid turbulence is much more complex, with high nonlinearities and multi-scale dynamics, which are difficult to predict by an approximate linear operator of DMD. This is the main reason the prediction accuracy reduces in the turbulence case. However, the satisfactory accuracy in turbulence reconstruction by using only one sensor still demonstrates the potential of the proposed DCDMD. Also note that, in this case, DCDMD has achieved over 50 times enhancement of computational efficiency compared to the conventional CSDMD (see Table. 5).

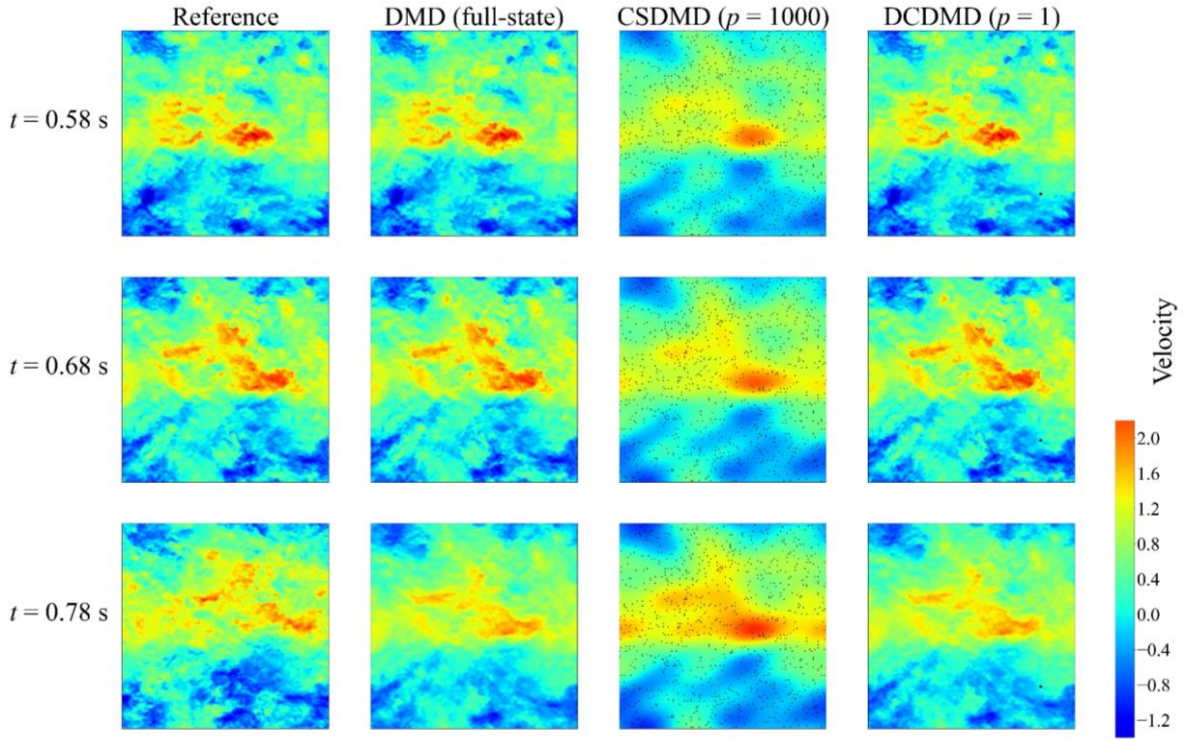

Fig. 20. Reconstructed velocity fields at 0.58, 0.68, and predicted velocity fields at 0.78 s:

(a) Reference; (b) DMD; (c) CSDMD; (d) DCDMD. Black dots in the third and fourth rows indicate randomly deployed sensors.

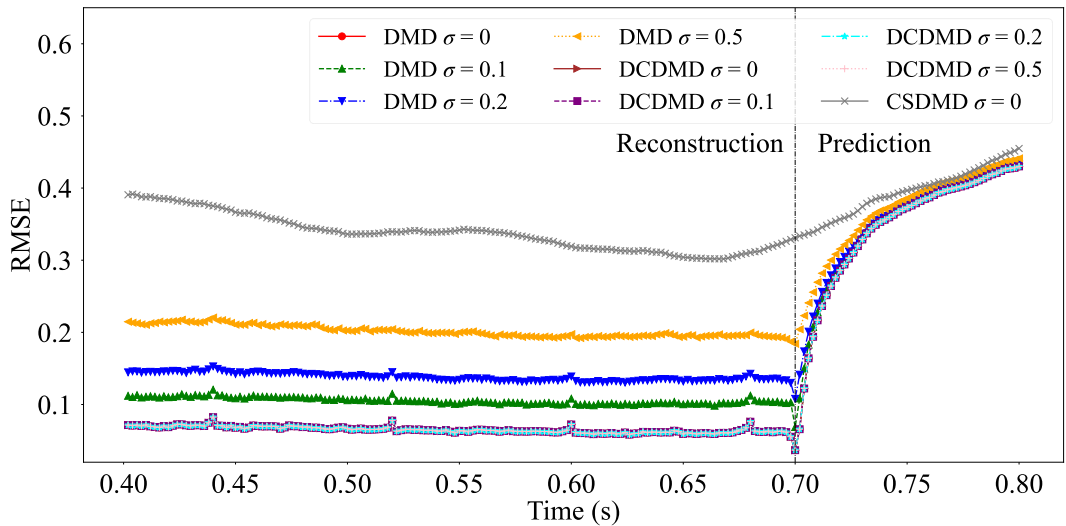

Fig. 21. Spatial-averaged RMSEs of DMD, CSDMD and DCDMD solutions subjected to different noise levels in the case of isotropic turbulence

Table. 4. Spatial-temporal averaged RMSEs of DMD, CSDMD and DCDMD solutions subjected to different noise levels in the case of isotropic turbulence

| | $\sigma$ | DMD (full field) | | CSDMD (1000 sensors) | | DCDMD (1 sensor) | |
| --- | --- | --- | --- | --- | --- | --- | --- |
| | | Reconstruction | Prediction | Reconstruction | Prediction | Reconstruction | Prediction |
| **RMSE** | 0 | $6.51\times10^{-2}$ | $3.56\times10^{-1}$ | $3.39\times10^{-1}$ | $3.96\times10^{-1}$ | $6.51\times10^{-2}$ | $3.56\times10^{-1}$ |
| **RMSE** | 0.1 | $1.05\times10^{-1}$ | $3.62\times10^{-1}$ | \ | \ | $6.52\times10^{-2}$ | $3.56\times10^{-1}$ |
| **RMSE** | 0.2 | $1.38\times10^{-1}$ | $3.63\times10^{-1}$ | \ | \ | $6.51\times10^{-2}$ | $3.56\times10^{-1}$ |
| **RMSE** | 0.5 | $2.02\times10^{-1}$ | $3.75\times10^{-1}$ | \ | \ | $6.51\times10^{-2}$ | $3.56\times10^{-1}$ |

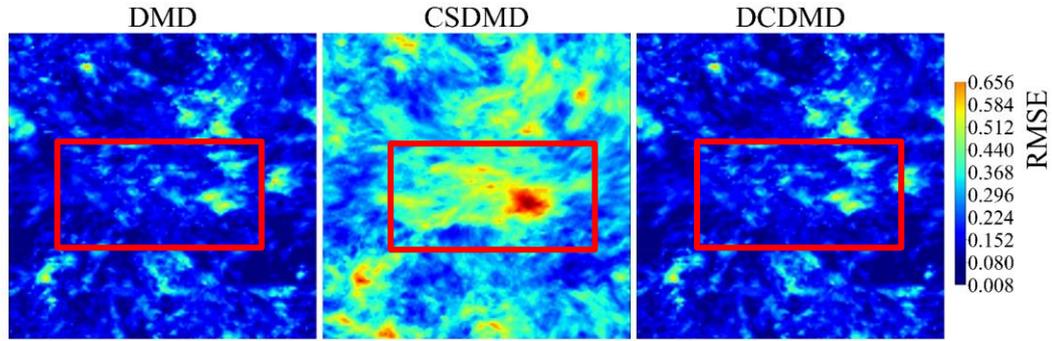

Fig. 22. Temporal-averaged RMSEs of the reconstructed velocities by DMD, CSDMD and DCDMD in the case of isotropic turbulence. Red rectangles represent the regions with high velocities.

Table. 5. Computational time of reconstruction/prediction of full-state flow fields by CSDMD and DCDMD in the case of isotropic turbulence

| Method | CSDMD | DCDMD |
| --- | --- | --- |
| **Computational time (s)** | 212.59 | 3.97 |

### 5.3.2.  Robustness assessment

By adding Gaussian white noises at different levels to the reference data, the velocity fields at $t = 0.78$ s reconstructed by DMD and DCDMD are presented in Fig. 23. With the noise ratios increasing to $\sigma = 0.5$, DMD predictions show growing deviations to the clean flow fields. In the entire reconstruction stage, the peak values of spatial-averaged RMSEs for $\sigma = 0$, 0.1, 0.2 and 0.5 are $6.51\times10^{-2}$, $1.05\times10^{-1}$, $1.38\times10^{-1}$ and $2.02\times10^{-1}$ (see Fig. 21 and Table. 4), respectively, corresponding to a three-fold increase of errors. On the other hand, the RMSEs of DCDMD solutions remain almost identical in cases with different noises, with the peak RMSE being around 0.06. In the prediction stage, although the RMSEs in various noise levels increase sharply for both methods, the differences among the DCDMD curves are smaller than those among DMD curves. The results at both the reconstruction and prediction stages demonstrate the insensitivity of DCDMD to measurement noises. Notably, the encouraging accuracy of DCDMD in managing the challenging turbulence problem by using data from one flexibly-deployed sensor is worth emphasizing.

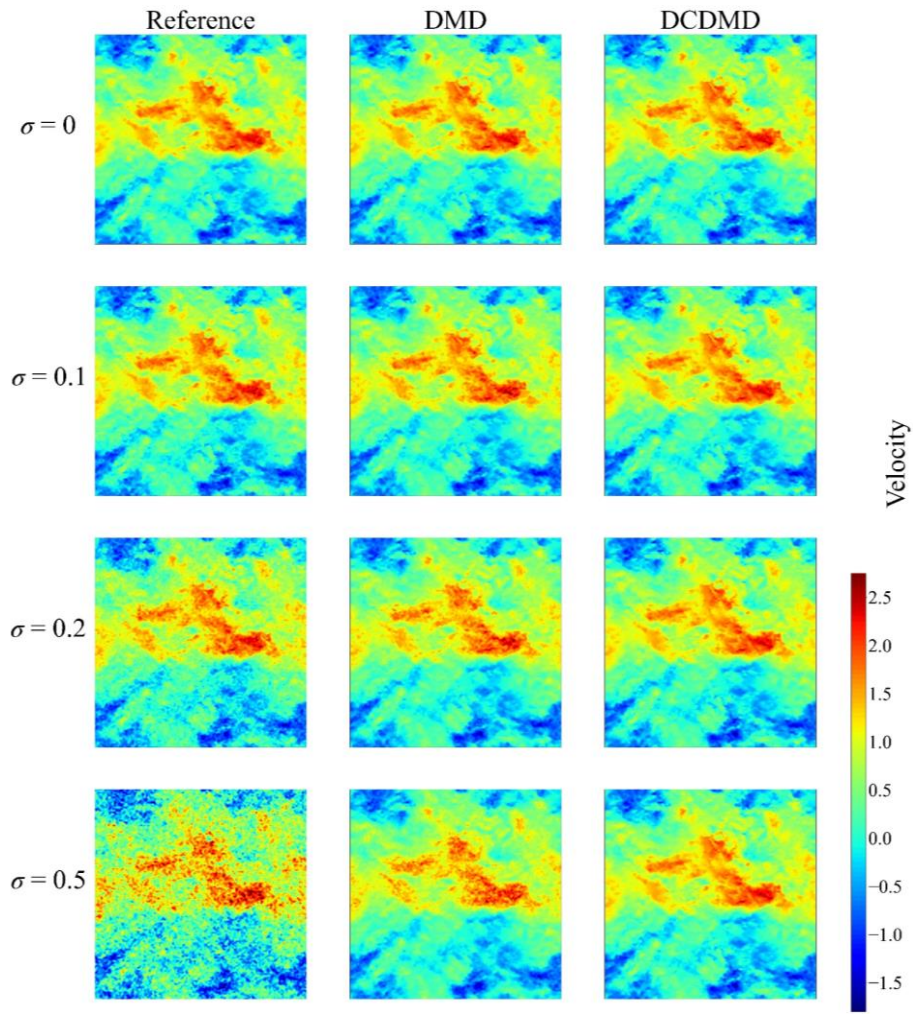

Fig. 23. Reconstructed velocity fields at 0.68 s with noise percentage of $\sigma = 0$, 0.1, 0.2 and 0.5: (a) Reference; (b) DMD; (c) DCDMD

## 6. Conclusions

In this study, the Deep-learning assisted Compressed-sensing Dynamic Mode Decomposition (DCDMD) is proposed for identifying coherent flow modes, as well as reconstructing and predicting full-state flow fields, from data at very few sampling locations. Advancing the conventional CSDMD, the novelties of the DCDMD model lie in two aspects: (1) defining a sparse/measurement matrix that allows flexible sensor deployments and requires a very small number of sensors; (2) building a DNN-based proxy (from prior knowledge or the so-called training data) that is capable of efficiently identifying DMD modes. These two advantageous features lead to the promising capacities of DCDMD in terms of accuracy, efficiency, robustness and practicality.

The performances of the proposed DCDMD model are demonstrated by three benchmark examples ranging from laminar to turbulent flows, i.e., flow past a cylinder, global sea surface temperature, and isotropic turbulence in a square periodic domain. In all three examples, DCDMD with a randomly deployed sensor produces very close solutions, in both reconstruction and prediction stages, to the traditional DMD utilizing full-state data. Noteworthy is that DCDMD only takes around 0.006% of the data (hence much few sensors) to obtain the same level of accuracy. Simulations with three noise-to-signal ratios, i.e., $\sigma = 0.1$, $0.2$ and $0.5$, are conducted. The results of different DCDMD simulations are almost identical, while those of DMD simulations show evident discrepancies, which demonstrate the insensitivity of DCDMD to the inevitable measurement noises, i.e., robustness.

As compared to CSDMD, DCDMD achieves more than 5.2 times higher accuracy in reconstruction and 1.1 times in prediction, by using only 0.1% of the number of sensors and costing around 1.9% of the computational time (numbers based on the turbulence example). Besides, the very few sensors (one adopted in the present study) utilized by DCDMD can be deployed flexibly according to technical difficulties and economy, rather than constrained by the stringent requirements on sensor numbers and sensor locations satisfying the random distribution. This is particularly helpful to the practicality of dealing with large-scale real-life problems.

Several points are worth emphasizing. Firstly, although DCDMD possesses high accuracy in full-state reconstruction in all three examples studied, the accuracy reduces in the prediction or extrapolation stage, especially in cases with complex fluid dynamics. This is attributed to the nature of DMD, which approximates the dynamic evolution of fluid flows as a linear process. Secondly, the proposed DCDMD has successfully reconstructed the highly-nonlinear flow features, such as the ENSO phenomenon in the global SST and the chaotic structures of fluid turbulence, as well as satisfactorily predicted the primary trends of these phenomena, yet, in general, the prediction accuracy needs improvements. Therefore, the nonlinear behaviors of the attractors to enhance the accuracy of prediction will be our future work. Another extension of the DCDMD model is to utilize the constraint-free feature of the sparse matrix for dealing with scenarios where sensor movements and/or data interruption exist.

# 7. Appendix

## 7.1. Numerical parameters in CSDMD

A critical parameter in CSDMD is the sparsity $Q$ in CoSaMP (which solves for DMD modes as a convex optimization problem). Due to the lack of prior knowledge of dynamics, there is currently no proper way with rigorous theoretical bases to determine $Q$. In the present study, we search for the optimal $Q$ in each CSDMD case by testing $Q$ values between 10 and 100 with an interval of 5. The number of maximum iterations, which is also key in CoSaMP, is set as 10 for balancing computational efficiency and accuracy. Another important parameter of CSDMD is the basis, in which the full-state information is sparse. The Fourier basis is selected in the present study due to its satisfaction of the sparsity condition and its suitability to the examples of the present study.

# 8. CRediT authorship contribution statement

**Jiaxin Wu**: Methodology, Formal analysis, Investigation, Data Curation, Writing - Original Draft, Writing - Review & Editing, Visualization, Project administration, Funding acquisition. **Dunhui Xiao**: Methodology, Writing - Review & Editing. **Min Luo**: Supervision, Methodology, Formal analysis, Investigation, Writing - Original Draft, Writing - Review & Editing, Project administration, Funding acquisition.

# 9. Acknowledgement

The authors are grateful to Dr. Jinlong Fu at Swansea University and Dr. Abbas Khayyer at Kyoto University for their insightful comments. The authors would like to express their appreciation to Prof. Kunihiko Taira at the Department of Mechanical and Aerospace Engineering, University of California, Prof. Tim Colonius at the California Institute of Technology, and Prof. Steve Brunton at the University of Washington, for kindly providing the data of the cylinder wake case. This research was supported by the start-up funding provided by Zhejiang University (to the corresponding author) and the Science Foundation of Donghai Laboratory (No. DH-2022KF0311).